\documentclass[aps,prb,twocolumn,floatfix,showpacs]{revtex4}
\usepackage{amsmath,amssymb,graphicx,bm}

\begin{document}



\title{ Ergodicity breaking in frustrated disordered systems: Replicas in mean-field spin-glass models}

\author{V. Jani\v{s}$^\ast$\thanks{$^\ast$Corresponding author. Email: janis@fzu.cz}, A. Kauch, and A. Kl\'\i\v{c} \\  {\em{Institute of Physics, Academy of Sciences of the Czech
  Republic, Na Slovance 2, CZ-18221 Praha 8, Czech Republic}}
}


\begin{abstract}
We discuss ergodicity breaking in frustrated disordered systems with no apparent broken symmetry of the Hamiltonian and present a way how to amend it in the low-temperature phase. We demonstrate this phenomenon on mean-field models of spin glasses. We use replicas of the spin variables to test thermodynamic homogeneity of ergodic equilibrium systems. We show that replica-symmetry breaking reflects ergodicity breaking and is used to restore an ergodic state. We then present explicit asymptotic solutions for the Ising, Potts and $p$-spin glasses. Each of the models shows a different low-temperature behavior and the way the replica symmetry and ergodicity are broken.    
\medskip
%
\end{abstract}
\maketitle

\section{Introduction}

Ergodicity is one of the most important properties of large statistical systems. We usually assume that macroscopic systems in equilibrium are ergodic.  Ergodicity, or quasi-ergodicity, means the phase trajectory comes arbitrarily close to any point of the space allowed by macroscopic constraints. The physically most important form of the ergodic theorem due to Birkhoff tells us that time average along the phase-space trajectory equals the statistical average over the whole phase volume.\cite{Birkhoff31} It means that the equilibrium state must span the whole available phase space. Although the foundations of statistical mechanics are based on ergodicity, lack of ergodicity is widespread in physical phenomena.\cite{Palmer82} Typical examples of ergodicity breaking are phase transitions with symmetry breaking. Broken ergodicity is sometimes used as a generalization of symmetry breaking.\cite{Bantilan81}  Although broken global symmetry is always accompanied by broken ergodicity, the converse does not hold. There are systems that break ergodicity without any apparent symmetry of the Hamiltonian being broken. The typical example is structural glass with numerous metastable states that prevent the macroscopic system from reaching the true equilibrium on experimental time scales.   

Broken ergodicity represents an obstruction in the application of fundamental thermodynamic laws. It hence must be recovered. When ergodicity is broken in a phase transition breaking a symmetry of the Hamiltonian, one introduces a symmetry breaking field into the Hamiltonian, being the Legendre conjugate to the extensive variable that is not conserved in the broken symmetry transformation in the low-temperature phase. The symmetry-breaking field allows one to circumvent the critical point of the symmetry-breaking phase transition and simultaneously restores ergodicity. Systems with no broken symmetry of the Hamiltonian at the phase transition, such as spin glasses or some quantum phase transitions, do not offer external symmetry-breaking fields and other techniques must be employed to find the proper portion of the phase space covered by the phase-space trajectory. 

In this paper we discuss the peculiarities of ergodicity breaking in phase transitions in frustrated disordered systems described by random non-local interaction without any directional preference in the phase space of the fundamental variables. We use mean-field spin-glass models with a Gaussian random spin-exchange and show how one can restore ergodicity via hierarchical replicating the spin variables of the original Hamiltonian. We present a general scheme of the real-replica method and apply it on mean-field Ising, Potts and $p$-spin glass models. The three generic models show different ways in which ergodicity is broken and we  demonstrate on them how the phase space of the equilibrium state can be constructed. We manifest that ergodicity breaking is equivalent to the replica-symmetry breaking.        

\section{Ergodicity and thermodynamic homogeneity}

Birkhoff ergodic theorem allows one to introduce equal \`a priori probability of allowed states in the phase space. It is, however, nontrivial to determine in macroscopic parameters which the allowed states indeed are. This is actually the most difficult task in constructing the proper phase space in statistical models. That is, to find out which points of the phase space are infinitesimally close to the trajectory of the many-body system extended to infinite times. Since we never solve the equation of motion of the statistical system, we have only static means to check validity of ergodicity. We then  test consequences of the ergodic hypothesis on the behavior of the equilibrium state. The most important consequence of ergodicity of statistical systems is the existence of the thermodynamic limit.    

The trajectory of the many-body system covers almost the whole allowed phase space. It means that the space covered by such trajectory does not depend on the initial state in non-chaotic systems. In ergodic systems then the thermodynamic limit does not depend on the specific form of the volume in which the macroscopic state is confined as well as on its surrounding environment. The macroscopic systems can either be isolated or embedded in a thermal bath. The thermodynamic equilibrium, the equilibrium state in the thermodynamic limit, is the same with vanishing relative statistical fluctuations. The thermodynamic equilibrium can then be reached by limiting any partial volume of the whole to infinity. The ergodic equilibrium state is homogeneous in the thermodynamic limit.        

Thermodynamic homogeneity is usually expressed via Euler's lemma\cite{Reichl80}
\begin{equation}\label{eq:Euler}
\alpha\ S(U,V,N,\ldots,X_i,\ldots) =  S(\alpha U, \alpha V,
\alpha N,\ldots,\alpha X_i,\ldots)
\end{equation}
telling us that entropy $S$ is an extensive variable and is a first-order homogeneous function  of all extensive variables, internal energy $U$, volume $V$ number of particles $N$, and model dependent other extensive variables $X_{i}$. As a consequence of Euler's lemma we obtain that thermodynamic equilibrium is attained as a one-parameter scaling limit where we have only one independent large scale, extensive variable, be either volume or number of particles, and the other extensive variables enter thermodynamic potentials as volume or particle densities insensitive to changes of the scaling variable.      

Thermodynamic homogeneity allows us to use scaling of the original phase space. Thermodynamic quantities remain unchanged if we arbitrarily rescale the phase space and then divide the resulting thermodynamic potential by the chosen scaling (geometric) factor. We can do that by scaling energy $E$ of the equilibrium state. If we use a scaling factor $\nu$, that can be an arbitrary positive number,  then the following identities hold for entropy $S(E)$ of the microcanonical and free energy $F(T)$ of the canonical ensemble  with energy $E$ and temperature $T$, respectively  
\begin{subequations}
\begin{align}\label{eq:Homogeneity}
S(E) &= k_B \ln \Gamma(E) =  \frac {k_B}\nu \ln \Gamma(E)^\nu =\ \frac{ k_B}
\nu \ln \Gamma(\nu E) \ , \\ F(T) &= -\ \frac {k_BT}\nu  
\ln\left[\text{Tr}\ e^{-\beta H}\right]^\nu = -\ \frac {k_BT}\nu  
\ln\left[\text{Tr}\ e^{-\beta \nu H}\right]\ ,
\end{align}
\end{subequations}
where we denoted $\Gamma(E)$ the phase-space volume of the isolated system with energy~$E$. 

The scaling of the phase space with an integer scaling factor $\nu$ can be simulated by replicating $\nu$-times the extensive variables. That is, we use instead of a single phase space $\nu$ replicas of the original space.  The reason to introduce replicas of the original variables is to extend the space of available states in the search for the allowed space in equilibrium. The replicas are independent when introduced. We use the replicated variables to study stability of the original system with respect to fluctuations in the thermal bath. To this purpose we break independence of the replica variables by switching
on a (homogeneous) infinitesimal interaction between the replicas that we
denote $\mu^{ab}$. We then add a small interacting part $\Delta H(\mu)=
\sum_{i} \sum_{a < b}^{\nu}  \mu^{ab} X_i^a X_i^b$ to the replicated 
Hamiltonian with dynamical variables $X_{i}$. The original system is then stable with respect to fluctuations in the bath, represented by the interaction with the replicated variables, if the linear response to perturbation $\mu$ is not broken.  If the linear response holds then the perturbed free energy per replica relaxes, after switching perturbation $\mu$ off, to the original one in the thermodynamic limit 
\begin{align}\label{eq:avFE}
  - \beta F_\nu(\mu)  &= \ \frac 1\nu
 \ln\mathrm{Tr}_{\nu}\exp\left\{-\beta\sum_{a=1}^{\nu} H^a -\beta
      \Delta H(\mu) \right\}\nonumber \\  &
\xrightarrow[\mu \to 0]{} \ln\mathrm{Tr}\exp\left\{-\beta H\right\} \ , 
\end{align}
where $\mathrm{Tr}_{\nu}$ refers to trace in the $\nu$-times replicated phase space.
If the linear response to the inter-replica interaction does not hold, the thermodynamic limit of the original system is not uniquely defined and depends on properties of the thermal bath. If there are no apparent physical fields breaking the symmetry of the Hamiltonian, the phase-space scaling represented by replicas of the dynamical variables introduces shadow or auxiliary 
symmetry-breaking fields, inter-replica interactions $\mu^{ab}>0$. They induce new order parameters in the response of the system to these fields that need not vanish in the
low-temperature phase, when the linear response breaks down. They
offer a way to disclose a degeneracy when the thermodynamic limit is not uniquely defined
by a single extensive scale and densities of the other extensive variables. The inter-replica interactions are not measurable and hence to restore the physical
situation we have to switch off these fields at the end. If the system is thermodynamically
homogeneous we must fulfill the following identity
\begin{equation}\label{eq:av-homogeneity}
\frac{d}{d\nu}\left[\lim_{\mu\to0} F_\nu(\mu) \right]\equiv 0 \ 
\end{equation}
for arbitrary $\nu$. This quantification of thermodynamic homogeneity, thermodynamic
independence of the scaling parameter $\nu$, will lead us in the
construction of a stable solution of mean-field spin glass models. To use equation~\eqref{eq:av-homogeneity} in the replica approach we will need to continue analytically the replica-dependent free energy to arbitrary positive scaling factors $\nu \in \mathbb R^{+}$. Specific assumptions on the symmetry of matrix $\mu^{ab}$ will have to be introduced. It is evident from Eq.~\eqref{eq:avFE} that the linear response to inter-replica interaction can be broken only if the replicas are mixed in the $\nu$-times replicated free energy $F_{\nu}$.

\section{Frustrated disordered systems - mean-field spin-glass models}
\label{sec:Disorder}

We present models on which the replica approach to the construction of the equilibrium state proves efficient. The linear response to a small inter-replica interaction may be broken only if replicas are intermingled in thermodynamic potentials. Mixing of replicas is achieved by randomness of an inter-particle interaction. We use lattice spin models with random spin-exchange to study replica mixing thermodynamic potentials. To simplify the reasoning we resort to mean-field models with no explicit spatial coherence. The mean-field approximations of lattice systems can either be introduced as models on fully connected graphs, models with long-range interaction, or as the limit to infinite spatial dimensions on hyper-cubic lattices. Thereby a scaling of the inter-site interactions is needed so that to keep the energy an extensive variable, linearly proportional to the volume as demanded by Eq.~\eqref{eq:Euler}.  

\subsection{Ising spin glass}

The simplest lattice spin system consists of spins with the lowest value $\hbar/2$ where only their projection to the easy axis enters interaction. We set in this paper $\hbar = 1, k_{B} = 1$.  The spins can then be treated classically  with projections $S_i= \pm 1$. The resulting Hamiltonian of the  Ising model reads
 \begin{equation}\label{eq:H-Ising}
 H \left[J,S\right]= \sum_{i<j} J_{ij} S_i S_j + h\sum_i S_i\ .
 \end{equation}
To obtain a glassy behavior we introduce  a randomness into spin exchange $J_{ij}$. We will explicitly consider only the mean-field version of this model with a Gaussian distribution of the spin-exchange.  The energy remains an extensive variable if we rescale the (long-range) interaction in the man-field limit as follows
$$
 N \left\langle J_{ij}\right\rangle_{av} = \sum_{j=1}^N J_{ij} = 0, \quad
  N\left\langle J_{ij}^2\right\rangle_{av} =  \sum_{j=1}^N J_{ij}^2 = J^2 \ ,
  $$
 where  $N$ is the number of lattice sites.  

\subsection{Potts glass}

The $p$-state Potts model is a generalization of the Ising model to $p> 2$ spin components.  The original formulation of Potts\cite{Potts52} with
 Hamiltonian $H_{p}=-\sum_{i<j}J_{ij}\delta_{n_{i},n_{j}}$ where
$n_{i}=0,\ldots, p-1$ is an admissible value of the $p$-state
model on the lattice site $\mathbf{R}_{i}$, is unsuitable for
practical calculations. The Potts Hamiltonian can, however, be represented via
interacting spins\cite{Wu82} %
\begin{equation}\label{eq:H-Potts}
  H_{P}\left[J,\mathbf{S}\right]=-\frac{1}{2}\sum_{i,j}J_{ij}\mathbf{S}_{i}\cdot \mathbf{S}_{j}  - \sum_{i}\mathbf{h}\cdot \mathbf{S}_{i}\ ,
\end{equation} 
where $\mathbf{S}_{i} = \{s_{i}^{1},\ldots s_{i}^{p-1} \}$ are Potts
vector variables taking values from a set of state vectors
$\{\mathbf{e}_{A}\}_{A=1}^{p}$. Functions on vectors $\mathbf{e}_{A}$
are in equilibrium fully defined through their scalar product
\begin{subequations}
\begin{align}\label{eq:Potts-restrict}
\sum_{A=1}^{p}e^{\alpha}_{A}=0\ , &\quad  
\sum_{A=1}^{p}e^{\alpha}_{A}e^{\beta}_{A}=p\ \delta^{\alpha \beta} \ , & e^{\alpha}_{A}e^{\alpha}_{B}=p\ \delta_{A B}-1
\end{align}
for  $\alpha\in\{1,...,p-1\}$. We use the Einstein summation convention for
repeating Greek indices of the vector components indicating a scalar
product of the Potts vectors. Using these properties we can construct an explicit representation of the Potts spin vectors
\begin{align}\label{eq:Potts-repre}
e^{\alpha}_{A}=\left\{
\begin{array}{ll}
   0 & A<\alpha \\
   \sqrt{\frac{p(p-\alpha)}{p+1-\alpha}}  & A=\alpha \\
   \frac{1}{\alpha-p}\sqrt{\frac{p(p-\alpha)}{p+1-\alpha}} &
   A>\alpha\ .
\end{array}
\right. 
\end{align}
\end{subequations}

Potts model shows a glassy behavior if we introduce randomness in spin exchange $J_{ij}$. In the mean-field limit we chose the following Gaussian distribution
\begin{equation}\label{eq:Potts-distrib}
  P(J_{ij})=\frac{1}{(2\pi
  J^{2}/N)^{1/2}}\exp{\frac{-(J_{ij}-J_{0})^{2}}{2J^{2}/N}}\ ,
\end{equation}
where $J_{0}=N^{-1}\sum_{j} J_{0j}$ is the averaged (ferromagnetic) interaction.

\subsection{$p$-spin glass}

Potts model is not the only interesting extension of the Ising model. Another generalization is the so-called $p$-spin model. It describes a system of Ising spins where the spin exchange connects a cluster of $p$ spins.    The Hamiltonian of such a model reads 
\begin{equation}\label{eq:H-p-spin}
H_{p}\left[J,S\right] = \sum_{1\le i_{1} < i_{2} < \ldots <i_{p}}J_{i_{1}i_{2}\ldots i_{p}}S_{i_{1}}S_{i_{2}}\ldots S_{i_{p}} \ .
\end{equation}
Randomness in the  spin exchange is again taken Gaussian so that the energy is an extensive variable in the large volume limit $N\to\infty$:\cite{Gross84}   
$$
P\left( J_{i_{1}i_{2}\ldots i_{p}}\right) = \sqrt{\frac{N^{p-1}}{\pi p!}} \exp \left\{ - \frac{ J_{i_{1}i_{2}\ldots i_{p}}\ ^{2}N^{p-1}}{J\ ^{2}p!}\right\} \ .
$$
This model is interesting in that we can analytically study the limit $p\to\infty$ for which we know an exact solution.\cite{Derrida80a,Derrida80b}


\section{Ergodicity and replica-symmetry breaking}
\label{sec:Ergodicity}

Randomness in the spin exchange causes mixing of replicas of the spin variables. Frustration prevents selection of easy axes, even if inhomogeneously distributed,  that could select kind of regular spin ordering. The spin-glass models do not provide us with apparent symmetry-breaking fields that could stabilize the low-temperature glassy phase.  There is, nevertheless, an ordered low-temperature phase with order parameters to be found. The determination of the proper phase space of homogeneous order parameters is the most difficult part in the search for the true equilibrium state in the spin-glass models. Replicas proved to be the only available means for reaching this goal.  

\subsection{Annealed and quenched disorder}

Randomness in the spin exchange, introduced in Sec.~\ref{sec:Disorder}, causes mixing of replicas and may lead to breaking of ergodicity manifested in a replica-symmetry breaking. We can, however, treat the disorder either dynamically or statically. That is, we can prepare the system so that the random configurations are thermally equilibrated, annealed disorder, and contribute to a single equilibrium state. In this situation we average over random configurations of the partition sum. If we have $\nu$ replicas of the spin variables we have to calculate the following configurationally averaged partition function   
\begin{subequations}\label{eq:FZ-av}\begin{multline}\label{eq:Z-av}
\left\langle Z^\nu_{N}\right\rangle_{av} = \int D[J]\mu[J]\prod\limits_{a=1}^{\nu} \prod\limits_{i=1}^{N}
d[\mathbf{S}^a_i]\rho[\mathbf{S}^a_i] \\ \times \exp\left\{ -\beta
\sum_{a=1}^\nu H[J,\mathbf{S}^a]\right\} \ ,
\end{multline}
where $\mu(J)$ and $\rho(\mathbf{S})$ are distribution functions for the spin exchange and spins, respectively. Or we can cool the macroscopic system down very fast so that there is not enough relaxation time  to reach the low-temperature equilibrium, quenched disorder, and we have to average thermodynamic potentials in the thermodynamic limit, e. g. free energy, 
\begin{multline}\label{eq:F-av}
-\beta \left\langle F^\nu_{N}\right\rangle_{av} = \int D[J]\mu[J] \ln \int\prod\limits_{a=1}^{\nu} \prod\limits_{i=1}^{N}
d[\mathbf{S}^a_i]\rho[\mathbf{S}^a_i] \\ \times \exp\left\{ -\beta
\sum_{a=1}^\nu H[J,\mathbf{S}^a]\right\} \ .
\end{multline}\end{subequations}
Spin glasses are assumed to be quenched, hence averaging over spin couplings from Eq.~\eqref{eq:F-av} is to be used.

From the mathematical point of view it is much more complicated to evaluate the quenched averaging, since logarithm is difficult to handle. Edwards and Anderson\cite{Edwards75} introduced the replica trick, $\ln x = \lim_{n\to 0} \left(x^{n} - 1 \right)/n $ to convert quenched into the annealed averaging. The replica trick was originally the reason to introduce replicas. Hence, to distinguish the replicas from the replica trick from those testing thermodynamic homogeneity, we call the former mathematical replicas  and the latter real replicas.  

At the end, there is no big difference in the annealed and quenched randomness when using replicated variables. In both cases we average a replicated partition sum. The only difference is that there is no limit to zero number of replicas for the annealed randomness. Instead, independence of the replica index is demanded. The annealed and quenched free energies can then be represented in  the $\nu$-times replicated phase space:   
\begin{subequations}
\begin{align}
\beta F_{an}&= - \frac 1{\nu}\lim_{N\to\infty}\ln \left\langle Z^\nu_{N}\right\rangle_{av}\ , \\ 
\beta F_{qu} & = - \lim_{\nu \to 0}\left[ \frac 1{\nu }\lim_{N\to\infty}\left( \left\langle Z^\nu_{N}\right\rangle_{av} - 1 \right)\right] \ .
\end{align}
\end{subequations}
What is common for both cases is that we have to continue analytically the free energy of the replicated system to arbitrary positive multiplication factor $\nu$. Either to test quantitatively thermodynamic homogeneity, Eq.~\eqref{eq:av-homogeneity}, or to perform the limit $\nu\to 0$ non-perturbatively.  

\subsection{Replica-symmetry breaking}

To find the equilibrium state for the spin glass models we will test validity of the linear-response to a small inter-replica interaction. Although ergodicity of the original model may be broken, we expect that it will be restored in an appropriately replicated space. The inter-replica interaction generates new order parameters $\chi^{ab} = \langle\langle S^a
S^b\rangle_T\rangle_{av} - q$ that will not vanish if ergodicity, replica-symmetry is broken. We denoted $q = \langle\langle S^a\rangle_T^2 \rangle_{av}$ that is the order parameter in the non-replicated (enlarged) space.  A $\nu$-times replicated free energy density of the Ising spin glass (quenched disorder) in the mean-field limit can be represented as\cite{Janis05} 
\begin{multline}\label{eq:FE-averaged-finite}
f_\nu = \frac{\beta J^2}{4} \left[\frac 1\nu\sum_{a\neq b}^\nu
\left\{\left(\chi^{ab}\right)^2 + 2 q\chi^{ab}\right\}
- (1 - q)^2\right]\\ -\frac
1{\beta\nu}\!\int\limits_{-\infty}^{\infty}\frac{d\eta}{\sqrt{2\pi}} \
e^{-\eta^2/2} \ln \text{Tr} \exp\left\{\beta^2J^2\sum_{a < b}^\nu
\chi^{ab}S^aS^b \right. \\ \left. + \beta \bar{h}\sum_{a=1}^\nu S^a\right\}\ ,
\end{multline}
where we denoted the fluctuating magnetic field $\bar{h} = h + \eta\sqrt{q}$.  Before we investigate validity of he linear response, we analytically continue the expression for the free energy to arbitrary (non-integer) positive replication indices $\nu \in \mathbb R^{+}$. Parisi found restrictions on the symmetry of the matrix of overlap susceptibilities $\chi_{ab}$ to make $f_{\nu}$ an analytic  function of $\nu$.\cite{Parisi80a,Parisi80b,Parisi80c} They are  
\begin{align}\label{eq:chi-hierarchy}
\chi^{aa} &= 0\ ,\quad \chi^{ab} = \chi^{ba}\ , \quad \sum_{c=1}^{\nu}\left(\chi^{ac} - \chi^{bc} \right) = 0 \ .
\end{align}
These restrictions reduce the number of independent overlap susceptibilities to $K\le \nu$. Let each independent susceptibility $\chi_{l}$ have multiplicity $m_{l}$. Parameters $m_{l}$ must be divisors of $\nu$ and obey a sum rule $\nu = 1 + \sum_{i=1}^{K}m_{i}$.  We generally denote $K$ the number of independent values of overlap susceptibility $\chi_{ab}$. An example of such a matrix for $\nu = 8$, $K= 3$  and $m_{l}= 2^{l-1}$ is illustrated in Fig.~\ref{fig:hierarchy}. 
\begin{figure}\begin{center}
$$\begin{pmatrix}
0& \chi_1&
\chi_2&\chi_2&
\chi_3&\chi_3&
\chi_3 &\chi_3 \\
\chi_1
&0&\chi_2&\chi_2&
\chi_3&\chi_3&
\chi_3 &\chi_3  \\
\chi_2&\chi_2&
{0}&\chi_1&
\chi_3&\chi_3&
\chi_3&\chi_3  \\
\chi_2&\chi_2&
\chi_1& {0}&
\chi_3&\chi_3&
\chi_3&\chi_3   \\
\chi_3&\chi_3
&\chi_3&\chi_3&
{0}& \chi_1&
\chi_2&\chi_2  \\
\chi_3&\chi_3&
\chi_3&\chi_3&
\chi_1& {0}&
\chi_2&\chi_2  \\
\chi_3&\chi_3&
\chi_3&\chi_3&
\chi_2&\chi_2&
{0}& \chi_1  \\
\chi_3&\chi_3&
\chi_3&\chi_3&
\chi_2&\chi_2&
\chi_1& {0}
\end{pmatrix}$$
\caption{Matrix of overlap susceptibilities $\chi_{ab}$ for $\nu=8$ and with three levels (hierarchies) of symmetry breaking $K=3$ exemplifying the structure allowing for analytic continuation to arbitrary positive $\nu$. \label{fig:hierarchy}}
\end{center}\end{figure}
It is now straightforward to calculate free energy for matrices $\chi_{ab}$ fulfilling criteria~\eqref{eq:chi-hierarchy}. We obtain 
\begin{subequations}
 \begin{multline}\label{eq:avfe-density}
    f_K(q,\{\chi\};\{m\}) = -\frac\beta 4 (1-q)^2 + \frac \beta 4
    \sum_{l=1}^K (m_l-m_{l-1})\chi_l \\ \times (2q + \chi_l) + \frac \beta2
    \chi_1 -\ \frac 1{\beta m_{K}} \int_{-\infty}^{\infty} \frac
    {d\eta}{\sqrt{2\pi}} e^{-\eta^2/2}
    \ln\left[\int_{-\infty}^{\infty} \right.  \\ \left. \frac {d\lambda_K}{\sqrt{2\pi}}
      e^{-\lambda_K^2/2}\left\{\dots \int_{-\infty}^{\infty} \frac
        {d\lambda_1}{\sqrt{2\pi}}
        e^{-\lambda_1^2/2}
        \left\{2\cosh   \left[\beta\left(h  \right.\right.\right.\right.\right. \\ \left.\left.\left.\left.\left. +
              \eta\sqrt{q} + \sum_{l=1}^{K}\lambda_l \sqrt{\chi_l -
                \chi_{l+1}}\right)\right]\right\}^{m_1}\ldots\right\}
      ^{m_K/m_{K-1}}\right] \end{multline}
with $\chi_{K+1}=0$ and $m_{0} = 1$. It may appear convenient to rewrite the free-energy density to another equivalent form
 \begin{multline}\label{eq:mf-avfe}
f_K(q;\Delta\chi_1,\ldots,\Delta \chi_K, m_1,\ldots,m_K) =
-\frac\beta 4 \left(1-q - \right. \\ \left. \sum_{l=1}^K\Delta\chi_l\right)^2 - \frac 1\beta
\ln 2  + \frac \beta 4 \sum_{l=1}^K
m_l\Delta\chi_l\left[2\left(q + \sum_{i=l}^{K}\Delta\chi_{i}\right)\right. \\ \left. -
\Delta\chi_l\right] - \frac 1\beta \int_{-\infty}^{\infty}
\mathcal{D}\eta\     \ln\   Z_K  
\end{multline}
\end{subequations}
where we ordered the parameters so that 
$\Delta\chi_l = \chi_l-\chi_{l+1}\ge\Delta\chi_{l+1}\ge 0$. We further used a short-hand notation for iterative partition functions
 $$ 
  Z_l =
\left[\int_{-\infty}^{\infty}\mathcal{D}\lambda_l\
Z_{l-1}^{m_l}\right]^{1/m_l} 
$$
with an abbreviation for a Gaussian differential
$\mathcal{D}\lambda \equiv {\rm d}\lambda\ e^{-\lambda^2/2}/\sqrt{2\pi}$. 
The initial partition function for the Ising spin glass is 
 $Z_0 =
\cosh\left[\beta\left(h + \eta\sqrt{q} + \sum_{l=1}^{K}\lambda_l
\sqrt{\Delta\chi_l} \right)\right]
$.
Free energy $f_{K}$ is an analytic function of geometric parameters $m_{i}$ that can now be arbitrary positive numbers. Equilibrium values of order parameters $\chi_{l}, m_{l}$ in representation~\eqref{eq:avfe-density} or $\Delta\chi_{l}, m_{l}$ from~\eqref{eq:mf-avfe} are determined from extremal points of the free energy functional. 

It is clear that complexity of the solution increases rapidly with the increasing number of different values $\Delta\chi_{i}$, that is, with number $K$ of replica hierarchies. We give here an example of the lowest replica-symmetry breaking free energy ($K=1$) 
\begin{multline}\label{eq:FE-1RSB}
f_{1}(q;\chi_{1},m_{1})  = -\frac \beta4(1-q - \chi_{1})^2  \\+ \frac \beta4m_{1}\chi_{1}(2 q + \chi_{1})   -\frac 1{\beta m_{1}}
\int_{-\infty}^{\infty}\mathcal{D}\eta \ln\int_{-\infty}^{\infty}
\mathcal{D}\lambda_{1} \\  \left\{2\cosh\left[\beta\left(h +
\eta\sqrt{q} +\lambda_{1}\sqrt{\chi_{1}}\right)\right]\right\}^{m_{1}}\ .
\end{multline}
It has three parameters, $q,\chi_{1}, m_{1}$ to be determined from stationarity of the free-energy functional from Eq.~\eqref{eq:FE-1RSB}. It represents a free energy with the first level of ergodicity breaking or replica-symmetry breaking (1RSB). Generally, free energy $f_{K}$ stands for ergodicity breaking on $K$ levels, $K$ generations of replicas ($K$RSB). 

Free energy $f_K(q;\Delta\chi_1,\ldots,\Delta \chi_K, m_1,\ldots,m_K) $ contains $2K + 1$ variational parameters, $q, \Delta\chi_{i},m_{i}$ for $i=1,2,\ldots K$ that are determined from stationarity of the free energy with respect to small fluctuations of these parameters. The replica construction introduced a new parameter $K$ that is not \`a priori determined. It can assume any integer value in the true equilibrium. The number of replica hierarchies is in this construction determined from stability conditions that restrict admissible solutions, stationarity points. A solution with $K$ levels is locally stable if it does not decay into a solution with $K+1$ hierarchies. A new order parameter in the next replica generation
$\Delta\chi$ may emerge so that $\Delta\chi_l > \Delta\chi >
\Delta\chi_{l+1}$ for arbitrary $l$. That is, the new order parameter may peel
off from $\Delta\chi_l$ and shifts the numeration of the order parameters
for $i>l$ in the existing $K$-level solution.  To guarantee that this does
not happen and that the averaged free energy depends on no more geometric
parameters than $m_1,\ldots,m_K$ we have to fulfill a set of $K+ 1$
generalized stability criteria that for our hierarchical solution
read for $l=0,1,\ldots,K$
\begin{multline}\label{eq:AT-hierarchical}
 \Lambda^{K}_{l} = 1  - \beta^2\left\langle\left\langle \left\langle 1 -
   t^2 + \right.\right.\right. \\ \left.\left.\left. \sum_{i=0}^{l} m_i \left(\langle
    t\rangle_{i-1}^2 - \langle t\rangle_i^2\right)\right\rangle_{l}^2
  \right\rangle_K\right\rangle_\eta\ge 0 
\end{multline}
with $m_{0}= 0$ and formally $\left\langle t\right\rangle_{-1}=0$. 
We introduced
short-hand notations $t \equiv \tanh\left[\beta\left(h + \eta\sqrt{q} + \sum_{l=1}^K
    \lambda_l\sqrt{\Delta\chi_l} \right)\right]$ and $\langle
t\rangle_l(\eta;\lambda_K,\ldots,\lambda_{l+1}) =
\langle\rho_l\ldots\langle\rho_1 t \rangle_{\lambda_1} \ldots
\rangle_{\lambda_l}$ with $\langle X(\lambda_l) \rangle_{\lambda_l} =
 \int_{-\infty}^{\infty}\mathcal{D}\lambda_l\
 X(\lambda_l)$ and 
$
\rho_l   = Z_{l-1}^{m_l}/\langle Z_{l-1}^{m_l}\rangle_{\lambda_l}
$. 
The lowest $K$ for which all stability conditions, Eq.~\eqref{eq:AT-hierarchical}, are fulfilled is an allowed equilibrium state. It need not, however, be the true equilibrium state, since the stability conditions test only local stability and cannot decide which of several extremal points is the true ground state. The stability conditions, Eq.~\eqref{eq:AT-hierarchical}, are necessary for the system to be thermodynamically homogeneous. They are, however, not sufficient to guarantee global thermodynamic homogeneity. Note that stability conditions from Eq.~\eqref{eq:AT-hierarchical} guarantee only local homogeneity, since they hold only for the optimal geometric parameters $m_{l}$ determined by stationarity equations. The global thermodynamic homogeneity, Eq.~\eqref{eq:av-homogeneity}, would demand $\Lambda^{K}_{K} \ge 0$ for arbitrary positive $m_{K}$.

\subsection{Continuous limit}

If free energy $f_K(q;\Delta\chi_1,\ldots,\Delta \chi_K, m_1,\ldots,m_K)$ is unstable for all finite $K$'s one has to perform the limit $K\to\infty$. Parisi derived a continuous version of the infinitely times replicated system by assuming 
$\Delta\chi_l = \Delta\chi/K \to dx$, and neglecting second and higher
powers of $\Delta\chi_l$ with the fixed index~$l$. This ansatz was based on the analysis of the first few hierarchical solutions of the Ising spin glass.\cite{Parisi80a,Parisi80b,Parisi80c} When performing the limit $K\to\infty$ in representation Eq.~\eqref{eq:mf-avfe} the free-energy functional can then be represented as\cite{Janis08} 
\begin{multline}\label{eq:FE-continuous} f(q,X; m(x)) = - \frac
  \beta4 (1 - q -X)^2 - \frac 1\beta \ln 2+ \frac {\beta }2
  \int_0^Xdx \\ \ m(x)\left[q + X - x \right] - \frac
  1\beta \left\langle g(X, h + \eta \sqrt{q})\right\rangle_\eta
\end{multline} %
where $\langle X(\eta)\rangle_\eta = \int_{-\infty}^\infty
\mathcal{D}\eta X(\eta)$. This free energy is only implicit since its interacting part $g(X,h)$ can be expressed only via an integral representation containing the solution itself 
\begin{subequations}
\begin{multline}\label{eq:g0}
  g(X,h) = 
  \mathbb T_x \exp\left\{\frac 12 \int_0^X dx
    \left[\partial_{\bar{h}}^2 
   \right.\right. \\ \left.\left. + m(x) g'(x;h + \bar{h})\partial_{\bar{h}}
    \right] \right\} g(h + \bar{h})\bigg|_{\bar{h}=0}\ ,
\end{multline}
 with $g(h) = \ln \left[\cosh\beta h\right]$.  The "time-ordering" operator $\mathbb T_x$ orders products of $x$-dependent non-commuting operators from left to right in a  $x$-decreasing succession. The exponent of the ordered exponential contains function $g'(x;h)= \partial g(x;h)/\partial h$ for $x\in [0,X]$ and is not known when $g(x;h)$ is not know on the whole definition interval. This derivative can also be expressed via an ordered exponential 
\begin{multline}\label{eq:g1}
  g'(X,h) =  \mathbb T_x \exp\left\{  \int_0^X dx
    \left[\frac 12 \partial_{\bar{h}}^2  \right.\right. \\ \left.\left. + m(x) g'(x;h +
      \bar{h})\partial_{\bar{h}} \right] \right\} g'(h +
  \bar{h})\bigg|_{\bar{h}=0}\ .
\end{multline}
\end{subequations}
It is an implicit but closed functional equation for the derivative $g^{\prime}(x;h)$ on interval $[0,X]$ for a given function $m(x)$. We have to know the full dependence of this function on parameter $x$ to evaluate the free energy with continuous replica-symmetry breaking. It is important to note that free energy $f(q,X; m(x))$ defines a thermodynamic theory independently of the replica method within which it was derived. It means that we can look for equilibrium states of spin-glass models without the necessity to go through instabilities of the discrete hierarchical replica-symmetry breaking solutions.    

Analogously we can perform the limit $K\to\infty$ with $m_{l} - m_{l+ 1} = \Delta m /K \to  dm$ in representation Eq.~\eqref{eq:avfe-density}.   We recall that $m_{l+1}< m_{l}$. We further on assume that $\chi_{1} - \chi_{2} > 0$ is not infinitesimally small.
The limiting free energy can then be represented as
\begin{multline}\label{eq:FE-mcontinuous} 
f\left(q, \chi_{1}, m_{1}, m_{0}; x(m)\right) = - \frac\beta4 (1 - q -\chi_{1})^2 \\ + \frac\beta 4 \left[m_{1} \left( q + \chi_{1}\right)^{2}  - m_{0}q^{2}\right] -  \frac {\beta}4
  \int_{m_{0}}^{m_{1}}dm \left[q + X_{0}(m)\right]^{2} \\ -  \frac 1\beta \left\langle g_{1}(m_{0}, h + \eta \sqrt{q})\right\rangle_\eta \ ,
\end{multline} %
where we denoted $X_{0}(m) = \int_{m_{0}}^{m} dm^{\prime}\, x(m^{\prime}) $, 
\begin{subequations}
\begin{multline}\label{eq:gm1}
  g_{1}(m_{0}, h) = 
  \overline{\mathbb T}_m \exp\left\{ \frac 12 \int_{m_{0}}^{m_{1}} dm\, x(m) 
    \left[ \partial_{\bar{h}}^2 \right.\right. \\ \left.\left.
   +  m g'_{1}(m;h + \bar{h})\partial_{\bar{h}}
    \right] \right\} g_1(h + \bar{h})\bigg|_{\bar{h}=0}\ .
\end{multline}
and $\overline{\mathbb T}_m $ is now $m$ increasing ordering. The input function is  
\begin{multline}\label{eq:g_1-Zero}
g_{1}(h) = \frac 1{m_{1}} \ln \int_{-\infty}^{\infty}\frac {d \phi}{\sqrt{2\pi}}  e^{-\phi^{2}/2} \left[ 2\cosh\left(\beta\left(h \right.\right.\right. \\ \left.\left.\left.+ \phi\sqrt{\chi_{1} - X_{0}(m_{1})}\right)\right) \right]^{m_{1}}\ .
\end{multline}\end{subequations}
This free energy better suits the case when the solution with a continuous RSB peels off from  a solution with one-level RSB or the two solutions  coexist. The space of the order parameters is restricted to an interval $0\le m_{1}\le 1$ and $0 \le X_{0}(m) \le \chi_{1}  \le 1$.

If free energy $f_{K}$ is not locally stable and at least one of stability conditions, Eq.~\eqref{eq:AT-hierarchical}, is broken for all $K$'s, it is a question whether the continuous limit $K\to\infty$   is locally stable. It can be shown that the continuous free energy $f(q,X; m(x))$ is marginally stable, fulfills the continuous version of stability conditions with equality.\cite{Janis08} It means that the continuous free energy does not break ergodicity, is always marginally ergodic in the whole spin-glass phase as the ferromagnetic model is only at the critical point.    

Both continuous free energies, Eq.~\eqref{eq:FE-continuous} and~\eqref{eq:FE-mcontinuous},  were derived as the limit of the number of replica hierarchies $K\to\infty$ where the distance between the neighboring hierarchies is infinitesimal, that is $\Delta\chi_{l} \propto K^{-1}$, as well as $\chi_{K}\propto K^{-1}$, and $\Delta\chi_{l}/\Delta m_{l}< \infty$ for each $l\le K$. Representation~\eqref{eq:FE-mcontinuous} is, however, more general, since it allows for a mixture of a discrete and continuous order parameters.   

The continuous free energies were derived for the Ising spin glass but they can be straightforwardly generalized to other spin-glass  models. The symmetry of the order parameters has to be adapted and the input single-site free energy $g$  or $g_{1}$ is to be appropriately modified.\cite{Janis11b,Janis13}   

\section{Equilibrium states in spin-glass models: Asymptotic solutions}

Spin-glass models experience a nontrivial ergodicity breaking that can be studied with hierarchical replications. The replica method does not tell us, however, how the true equilibrium looks like. Whether finite number of replica hierarchies are enough to restore ergodicity or it is the marginally ergodic continuous limit that represents the equilibrium state. The problem with the mean-field theories of spin glasses is that we are unable to find full solutions in most models beyond the first level of ergodicity breaking, 1RSB free energy, Eq.~\eqref{eq:FE-1RSB}. The only way to resolve the stationarity equations of the full mean-field models is to use an asymptotic expansion in small order parameters below the transition point to the glassy phase, if they exist. That is, when the transition is continuous.  We now discus the three spin-glass models defined in Sec.~\ref{sec:Disorder}. Each of the models shows a different behavior in the glassy phase.    

\subsection{Ising spin glass}

Mean-field Ising spin glass is paradigm for the theory of spin glasses. It is called Sherrington-Kirkpatrick model.\cite{Sherrington75} It is this model for which Parisi derived a free energy with a continuous replica-symmetry breaking.\cite{Parisi80a,Parisi80b,Parisi80c} It was also later proved that the hierarchical scheme of replica-symmetry breaking covers the exact equilibrium state.\cite{Guerra03,Talagrand06}  The rigorous proof does not, however, tell us whether the equilibrium state is described only by a finite number of replica hierarchies  or a continuous limit is needed. Only a few years ago we resolved the hierarchical free energy $f_K(q;\Delta\chi_1,\ldots,\Delta \chi_K, m_1,\ldots,m_K)$ for arbitrary $K$ via the asymptotic expansion below the transition temperature in a small parameter $\theta = 1 - T/T_{c}$.\cite{Janis06} Only this asymptotic solution was able to resolve the question of the structure of the equilibrium state. We found the leading order of the order parameters 
\begin{subequations}\label{eq:order-parameters}
\begin{align}\label{eq:chi_i}
  \Delta\chi_l^K &\doteq \frac 2{2 K + 1}\ \theta\ ,\\ \label{eq:m_i}
  m_l^K &\doteq \frac {4 (K - l + 1)}{2 K + 1}\ \theta \ ,\\ \label{eq:q}
  q^K &\doteq \frac 1{2 K + 1}\ \theta
\end{align}
that clearly indicate that the limit $K\to\infty$ leads to the Parisi continuous replica-symmetry breaking.  Since each solution with a finite number of replica generations is unstable
\begin{equation}\label{eq:Lambda-solution}
 \Lambda_l^{K} = -\frac  43 \ \frac {\theta^2}{(2 K + 1)^2} < 0\ ,
\end{equation}
\end{subequations}
it is explicitly proved that the Parisi solution is the equilibrium state in the glassy phase of the Sherrington-Kirkpatrick model. 
Other physical quantities in the asymptotic limit are the Edwards-Anderson parameter defined as $Q^{K}= q + \sum_{l=1}^{K}\Delta \chi_{l}^{K}$ 
\begin{subequations}
\begin{align}\label{eq:Qprime}
  Q^K &\doteq \theta + \frac {12 K(K + 1) + 1}{3(2 K +1)^2}\ \theta^2\  ,
\end{align}
local spin susceptibility
\begin{equation}\label{eq:thermal-susceptibility}
\chi_T = \beta \left(1 - Q^{K} + \sum_{l=1}^K m_l\Delta\chi_l\right) \doteq 1 -
\frac{\theta^2} {3(2K + 1)^2}\ . 
\end{equation}
and the free-energy difference to the paramagnetic state, 
\begin{equation}\label{eq:FE-better-asymptotic}
\Delta f\doteq\left(\frac{1}{6}\theta^3+\frac{7}{24}
\theta^4+\frac{29}{120}\theta^5\right)-\frac{1}{360} \theta^5
   \left(\frac{1}{K}\right)^4 \ .
\end{equation}
\end{subequations}
Differences between different levels of RSB manifest themselves  in free energy first in fifth order.

\subsection{Potts glass}

The Potts model with $p$ states reduces to Ising for $p=2$, but differs from it for $p>2$ in that it breaks the spin-reflection symmetry. This property was used to argue that the Parisi scheme fails.\cite{Goldbart85} It had been long believed that it is the one-level replica-symmetry breaking that determines the equilibrium state below the transition temperature.\cite{Cwilich89} The Potts glass displays a discontinuous transition into the replica-symmetry broken state for $p>4$.\cite{Gross85} Discontinuous transitions do not allow us to use an asymptotic expansion in a small parameter below the transition temperature. It is, nevertheless, possible to test the ordered phase of the Potts glass for $2< p< 4$. We did it in Refs.~\onlinecite{Janis11a,Janis11b} and found an unexpected behavior.    

Studying the discrete replica-symmetry breaking we found two 1RSB solutions with the same geometric parameter  
\begin{align}
m &\doteq \frac{p-2}{2}+\frac{36-12 p +p^2}{8(4-p)}\theta\ .
\end{align}
One non-trivial 1RSB solution then leads to order parameters
\begin{subequations}\label{eq:1RSB-Sol1}
\begin{align}
q^{(1)}&\doteq 0\ ,\\
\Delta\chi^{(1)}&\doteq \frac{2}{4-p}\theta + \frac{228-96 p +p^2}{6 (4-p)^3}\theta^2
\end{align}
\end{subequations}
while the second one has both parameters nonzero
\begin{subequations}\label{eq:1RSB-Sol2}
\begin{align}\label{eq:1RSB-Solq2}
q^{(2)}&\doteq \frac{-12+24 p -7 p^2}{3(4-p)^2 (p-2)}\theta^2 \ ,\\
\Delta\chi^{(2)}&\doteq \frac{2}{4-p}\theta - \frac{360-204 p -6 p^2+13 p^3}{6 (4-p)^3 (p-2)}\theta^2 \ . \label{eq:1RSB-Solchi2}
\end{align}\end{subequations}
We can see that the asymptotic expansion with small parameters $q$ and $\Delta\chi$ breaks down already at $p=4$ above which we expect a discontinuous transition from the paramagnetic to a 1RSB state at $T_{0} > T_{c}=1$. Note that a transition to the replica-symmetric solution $q> 0,\Delta\chi =0$ is continuous up to $p=6$. 

The 1RSB solution has a higher free energy than the replica-symmetric one. The difference is of order $\theta^{3}$,  
\begin{equation}
f_{1RSB} - f_{RS} \doteq \frac {(p-2)^{2}(p-1) \theta^{3}}{3(4-p)(6-p)^{2}}\ .
\end{equation}
The two stationary states of the 1RSB free energy behave differently as a function of parameter $p$. The former solution is physical for all values of $p$ unlike the latter that becomes unphysical for $p>p^{*}\approx 2.82$ where $q^{(2)}$ from Eq.~\eqref{eq:1RSB-Solq2} turns negative. It is also the region of the parameter $p$ where the first solution is locally stable as can be seen from the stability function      
\begin{equation} \label{eq:1RSB-instability1}
\Lambda^{(0)}_{1} 
\doteq\frac{\theta^{2}(p - 1)}{6(4 - p)^{2}} \left(
    7p^{2} - 24 p + 12\right)  > 0\ .
\end{equation} 
in this region. That is why the solution with $q=0$ was assumed to be the true equilibrium and a solution with a continuous replica-symmetry breaking had not been expected to exist. We, however, found that there is a Parisi-like solution even in the region of stability of the solution from Eq.~\eqref{eq:1RSB-Sol1}. The second 1RSB solution is unstable and decays to solutions with higher numbers of replica hierarchies as 
\begin{subequations}
\begin{align}\label{eq:KRSB-q_As}
q^{K}&\doteq -\frac{1}{3K^{2}}\frac{12-24p+7p^{2}}{(4-p)^{2}(p-2)}\theta^{2}\ ,\\
\label{eq:KRSB-Dchi_As}
\Delta\chi_{l}^{K}&\doteq \frac{1}{K}\frac{2}{(4-p)}\theta \ ,
\\  \label{sol:chi}
m_{l}^{K}& \doteq\frac{p-2}{2}+
\frac{2}{4-p}{\left[3+\frac{3}{2}p-p^{2}
 \right.} \nonumber \\ &\qquad{\left. +\left(3-6p+\frac{7}{4}p^{2}\right)\frac{2l-1}{2K}\right]\theta}\ .
\end{align}
\end{subequations}
We can see that the $K$RSB solution behaves unphysically in the same way as the second 1RSB solution does. The averaged square of the local magnetization is negative for $p > p^{*}$ where the first 1RSB solution is locally stable. Negativity of $q$ means that local magnetizations are  imaginary and the solution is unphysical. This deficiency, however, decreases with the increasing number of spin hierarchies and disappears in the limit $K \to \infty$. It means that the resulting solution with a continuous replica-symmetry breaking shows no unphysical behavior. It is analogous to negativity of entropy in the low-temperature solutions of $K$RSB approximations of the Sherrington-Kirkpatrick model.  

\begin{figure}\begin{center}
\includegraphics[width=9cm]{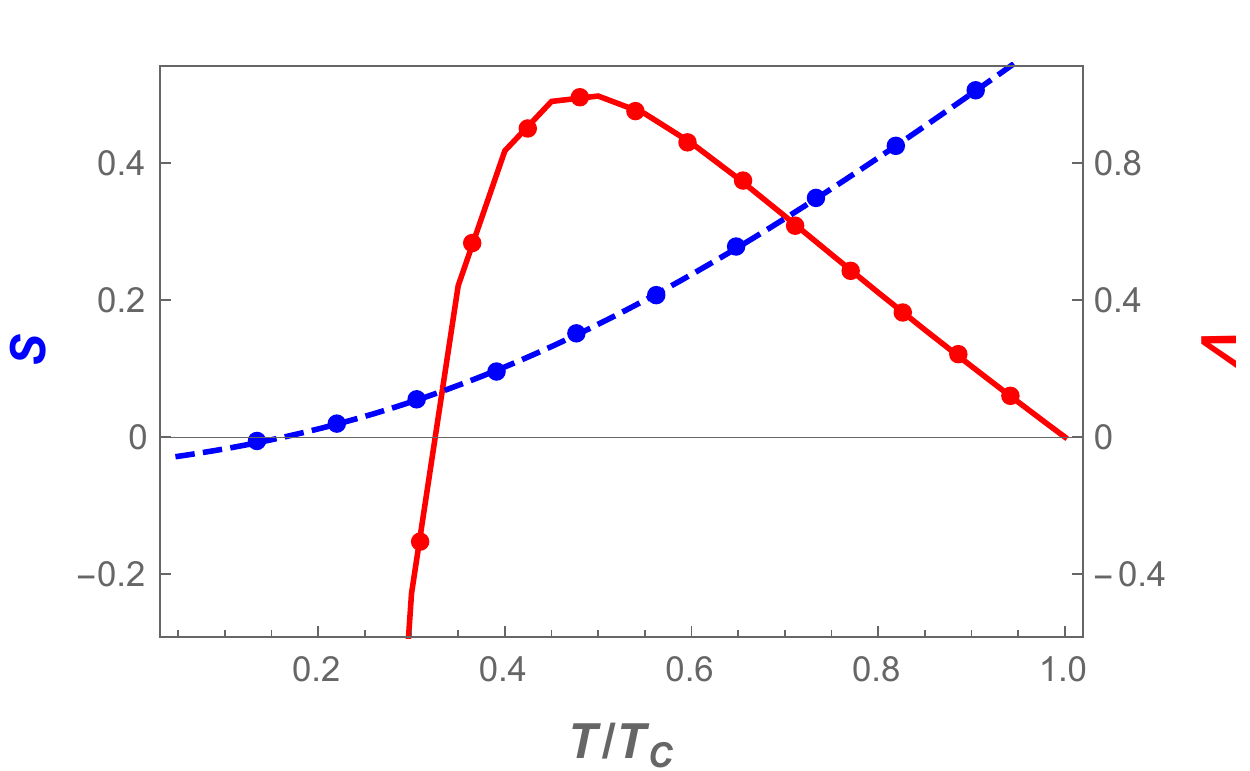}
\caption{Entropy  $S$ (left scale, dashed line)  and local stability $\Lambda$ (right scale, solid line) of the 1RSB solution from Eq.~\eqref{eq:1RSB-Sol1} of the $3$-state Potts glass. The solution becomes locally unstable at $\approx 0.33 T_{c}$ and entropy negative at $T\approx 0.16 T_{c}$.\label{fig:1RSB-Potts}}\end{center}
\end{figure}

Potts glass hence shows a degeneracy for $p^{*}< p< 4$ with a marginally stable solution continuously breaking  the replica symmetry and a locally stable one-level replica-symmetry breaking. To decide which one is the true equilibrium state one has to compare free energies. The difference of the continuous free energy $f_{c}$ and that of the $K$RSB solution is
\begin{subequations}
\begin{equation}\label{eq:fdiff} \beta (f_{c}-f_{KRSB})\doteq
  \frac{(p-1) (p (7 p-24)+12)^2 \theta ^5}{720 K^{4}(4-p)^5} 
\end{equation} 
and and that of the replica-symmetric one reads
\begin{equation}
  \beta(f_{c } - f_{RS}) \doteq \frac{(p - 1) (p - 2)^{2}\theta^{3}}{3(4 - p)(6
    - p)^{2}}\ .
\end{equation} 
\end{subequations}
We see that the solution with the continuous RSB has the highest free energy as the true equilibrium state should have for geometric factors $m<1$. In this situation entropy reaches minimum and free energy maximum in the phase space of the order parameters. The locally stable 1RSB solution becomes unstable at lower temperatures and entropy turns negative at very low temperatures as demonstrated on the $3$-state model in Fig.~\ref{fig:1RSB-Potts}. 
This leads us to the conclusion that the Parisi solution with a continuous replica-symmetry breaking represents the equilibrium state for the Potts glass with $p<4$.    

\subsection{$p$-spin glass}

The spin model generalized to random interactions connecting $p$ spins, $p$-spin glass has been used to simulate the dynamical transition in real glasses.\cite{Kirkpatrick87a,Kirkpatrick87b} This model, analogously to the Potts glass, generalizes the Ising spin glass to $p>2$ and allows one to study the behavior of the equilibrium state as a function of parameter $p$. In particular, the limit $p\to\infty$ is accessible\cite{Gross84} and is exactly solvable. It coincides with the random energy model of Derrida.\cite{Derrida80a,Derrida80b} For this reason the $p$-spin glass was also intended to be used to study and understand the genesis of the Parisi free energy when studying the asymptotic limit $p\to \infty$.

To cover both the boundary solutions $p=2$ and $p=\infty$ we have to mix up the one-level RSB scheme and the Parisi continuous RSB. Such free energy density of the mean-field $p$-spin glass reads
\begin{widetext}
\begin{multline}\label{eq:FE-general}
f^{(p)}_{T}(q, \chi_{1},\mu_{1},\mu_{0}; x(\mu)) =  - \frac{\beta}{4}\left[ 1 - p\left(q + \chi_{1}\right)  + (p-1) \left(q + \chi_{1}\right)^{p/(p-1)} \right]   \\ 
+ \frac{p-1}{4} \left[\mu_{1}\left(q + \chi_{1})\right)^{p/(p-1)} - \mu_{0}q^{p/(p-1)}\right]  - \frac {p - 1} 4 \int_{\mu_{0}}^{\mu_{1}} d \mu \left[q +  X_{0}(\mu) \right]^{p/(p-1)}  -  \left\langle  g_{1}\left(\mu_{0},h + \eta \sqrt{pq/2}\right) \right\rangle_{\eta}\ ,
\end{multline}
\end{widetext}
with 
\begin{subequations}
\begin{multline}\label{eq:g0x}
  g_{1}(\mu_{1},h)  = \overline{\mathbb T}_{\mu} \exp\left\{ \frac p 4 \int_{\mu_{0}}^{\mu_{1}} d \mu\ x(\mu) \left[\partial_{\bar{h}}^2 \right.\right. \\ \left. \left. 
 + \ \mu g_{1}^{\prime}(\mu,h + \bar{h})\partial_{\bar{h}}
    \right] \right . \bigg\} g_{1}(h + \bar{h})\bigg|_{\bar{h}=0} \ ,
\end{multline}
where  $X_{0}(\mu) =  \int_{\mu_{0}}^{\mu}d\mu' x(\mu')$. The generating free energy is 
\begin{multline}\label{eq:g_{mu}-Zero}
g_{1}(h) = \frac 1\mu_{1} \ln \int_{-\infty}^{\infty}\frac {d \phi}{\sqrt{2\pi}}  e^{-\phi^{2}/2} \left[ 2\cosh\left(\beta\left(h  \right.\right.\right. \\ \left.\left. \left.  + \phi\sqrt{p(\chi_{1} - X_{0}(\mu_{1})/2}\right)\right) \right]^{\mu_{1}/\beta}\ .
\end{multline}
\end{subequations}
We rescaled the function $m \to \mu = \beta m$.  If $M= \mu_{0}$, free energy $f^{(p)}_{T}$ reduces to the 1RSB approximation. On the other hand if $\chi_{1}=0$, or $\mu_{1}= \beta$ free energy  $f^{(p)}$ coincides with that of the Parisi solution with a continuous replica-symmetry breaking. 

\begin{figure*}\begin{center}
\includegraphics[width=12cm]{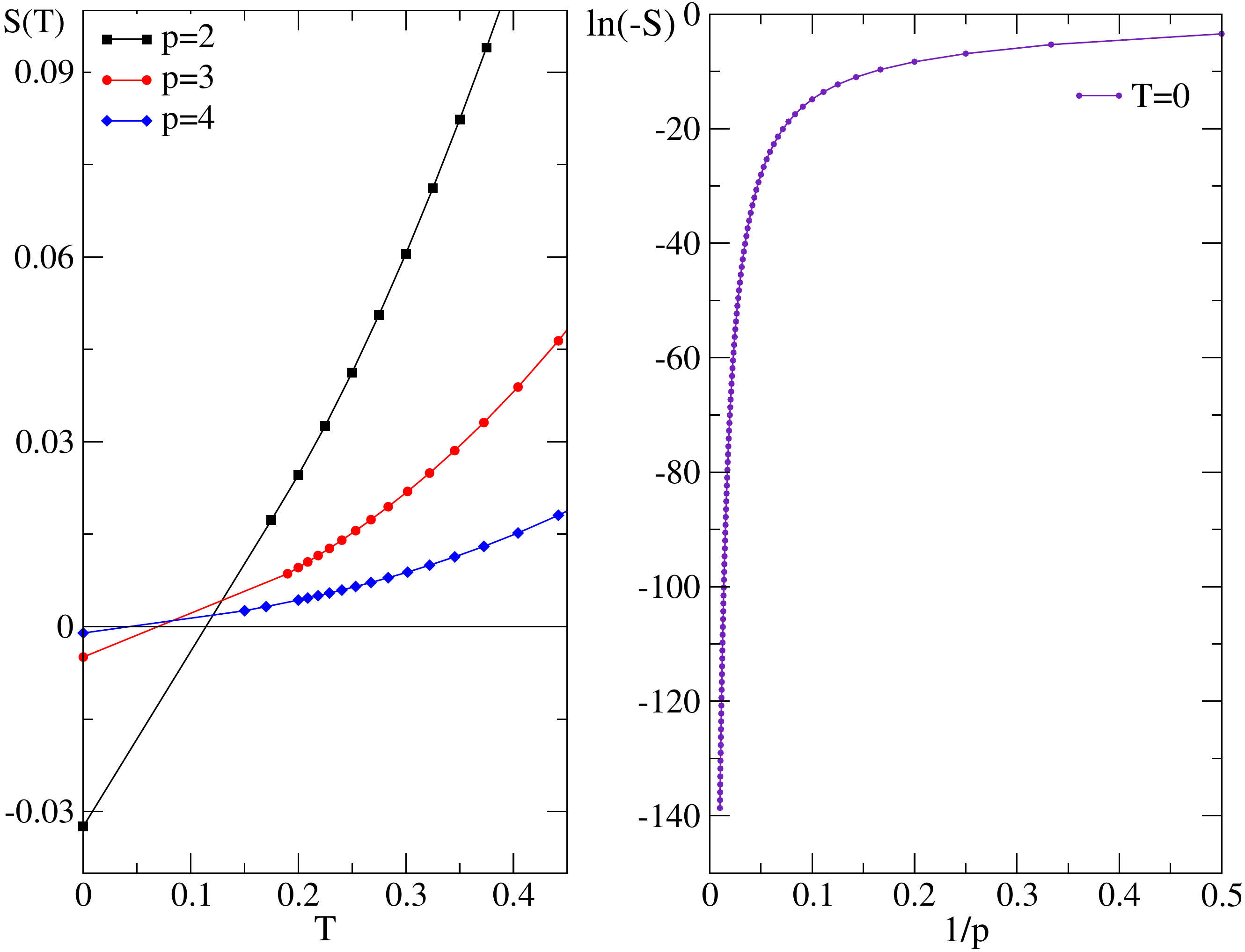}
\caption{Entropy of the 1RSB solution of the $p$-spin glass. Temperature dependence for  different values of $p$ (left panel). Logarithm of the negative part of entropy at zero temperature as a function of $1/p$  (right panel). \label{fig:1RSB-entropy}}\end{center}
\end{figure*}

The $p$-spin glass can not only be used to investigate analytically the $p\to\infty$ limit but also the $T\to 0$ limit. In this limit simple solutions of mean-field models lead to negative entropy. It is easy to calculate the zero-temperature entropy in the 1RSB solution. We obtain
\begin{align}\label{eq:S-Tlow}
S_{0}(h) &\propto  - \frac{p(p-1)}{8} \left[\frac{\exp\{- \mu_{1}^{2}p\chi_{1}/4\}}{\sqrt{\pi p\chi_{1}}}\  \frac{\exp\{- h^{2}/p\chi_{1}\}}{2CH_{\mu}(h)}\right]^{2}  \ , 
\end{align}
where we used the following notation
\begin{subequations}
\begin{align}\label{eq:E-CH}
2CH_{\mu}(h) & = e^{\mu_{1}h} E_{\mu}^{(p)}(- h) + e^{-\mu_{1}h} E_{\mu}^{(p)}(h)  \ ,
\\ \label{eq:E-mu}
E_{\mu}^{(p)}(h) &= \int_{h/\sqrt{p\chi_{1}/2}}^{\infty} \frac{d \phi}{\sqrt{2 \pi}} e^{-\left(\phi - \mu_{1}\sqrt{p\chi_{1}/2}\right)^{2}/2} \ .
\end{align}
\end{subequations}
Negativity of the low-temperature entropy indicates that 1RSB cannot produce a stable ground state for arbitrary $p<\infty$. Negativity of entropy decreases with increasing $p$, see Fig.~\ref{fig:1RSB-entropy}, but only if a condition $\mu_{1}^{2} p \chi_{1} = \infty$ is fulfilled the 1RSB solution ($\mu_{1}>0$) leads to zero entropy at zero temperature. Nonnegative entropy is a necessary condition for physical consistency of the low-temperature solution. It then means that the low-temperature equilibrium state for $p<\infty$ must contain the Parisi continuous order-parameter function $x(\mu)$ with $\beta \ge \mu_{1}\ge M >\mu_{0}$. It can also be seen from the asymptotic free energy for $p\to\infty$ that reads
\begin{widetext}
\begin{multline}\label{eq:FE-infty}
f^{(p\to\infty)}_{T}(q,\chi_{1},\mu_{1}) =  - \frac{1}{4T} \left[ 1 - \left(q + \chi_{1}\right)\left( 1 - \ln \left(q + \chi_{1}\right)\right)\right] - \frac 1{\mu_{1}}\ln \left[2\cosh(\mu_{1}h)\right]  \\  - \frac{\mu_{1}}{4} \left[ \chi_{1} - \left(q + \chi_{1}\right) \ln \left(q + \chi_{1}\right)\right]   -  \frac{\mu_{1}q}{4}\left[ \ln q  + p \left(1 - \tanh^{2}(\mu_{1}h)\right) \right]  \ ,
 \end{multline}\end{widetext}
giving the leading-order solution for the variational parameters $\chi_{1},\mu_{1}$ and $q$.  The first two parameters are of order one while the latter is exponentially small for large~$p$,  
\begin{subequations}\label{eq:stationary-asymptotic}
\begin{align}
\chi_{1} &=1 - q \ ,\\ 
q &= \exp\{- p(1 - \tanh^{2}(\mu_{1}h) )\}\ , \\
 \mu_{1} &= 2 \sqrt{\ln \left[2\cosh(\mu_{1}h)\right]  - h\tanh(\mu_{1}h)} \ .
\end{align}
\end{subequations}
The above nontrivial solution holds only if $\beta > 2 \sqrt{\ln \left[2\cosh(\beta h)\right]  - h\tanh(\beta h)}$ (low-temperature phase), otherwise $\mu_{1}= \beta$  and $\chi_{1} + q=0$ (high-temperature phase). To derive an equation for the order-parameter function $x(\mu)$ one needs to include the next-to-leading order contributions. To go beyond the leading asymptotic order one can use the Landau-type theory for the order-parameter function developed in Ref.~\onlinecite{Janis13}. Note that the asymptotic solution from Eq.~\eqref{eq:stationary-asymptotic} with $\mu(x) = 0$ suffers from a negative entropy as can be seen from Eq.~\eqref{eq:S-Tlow} and is plotted in Fig.~\ref{fig:1RSB-entropy}.  Note that the transition to the ordered phase in the $p$-spin glass is discontinuous and hence, an asymptotic expansion below the transition temperature is not applicable. Only an asymptotic expansion $p\to \infty$ makes sense. 

\section{Conclusions}

We studied ergodicity breaking that is not accompanied by any broken symmetry of the Hamiltonian.  Equilibrium thermodynamics and the equilibrium cannot be found then by standard methods, since ergodicity cannot be straightforwardly restored. When no symmetry of the Hamiltonian is broken we do not have direct means, physical macroscopically controllable external fields, with which we could remove dependence of the thermodynamic limit on properties of the thermal bath. Broken ergodicity impedes the existence of a uniquely defined thermodynamic limit that depends on the behavior of the thermal bath. The system is not thermodynamically homogeneous.   

We used replications of the phase space of the dynamical variables to simulate the impact of the thermal bath. We introduced a small inter-replica interaction and looked at the linear response to it. This scheme was hierarchically used to test and restore at least local thermodynamic homogeneity. The principal step in this procedure was to select an adequate symmetry-breaking of the replicated variables so that to make thermodynamic potentials analytic functions of the originally integer replication index. Only then it is possible to test and restore thermodynamic homogeneity and ergodicity. 

We explicitly demonstrated this construction on mean-field models of spin glasses. Randomness in the spin-exchange in these models makes them frustrated and no regular long-range order is established in equilibrium. Ergodicity is broken in the whole low-temperature, spin-glass phase and the thermodynamic limit is not uniquely defined. Replica method allows one to restore ergodicity in hierarchical steps by breaking successively the replica symmetry or independence of the replicated spaces.

We applied the real replicas on mean-field Ising, $p$-state Potts and $p$-spin glass models and calculated their asymptotic solutions below the transition temperature to the glassy phase. While the Ising spin glass is known to have continuously broken replica symmetry in the equilibrium state, the Potts and $p$-spin glasses allow for locally stable solutions with a one-level discrete replica-symmetry breaking. The continuous RSB and 1RSB coexist in the $p$-state Potts model, while the $p$-spin glass reduces to a 1RSB solution in the limit $p\to\infty$. In both cases for $p<\infty$ the 1RSB state leads to negative entropy at very low temperatures and it seems that the ultimate equilibrium state for the mean-field spin-glass models  breaks the replica symmetry in a continuos form as suggested by Parisi in the Ising model. Our analysis indicates that a continuous RSB is indispensable to prevent entropy to become negative at zero temperature.  Spin reflection symmetry is hence not substantial for the existence of a solution with a continuous RSB.


\begin{thebibliography}{12}

\bibitem{Birkhoff31} Birkhoff GD. Proof of the ergodic theorem.  Proc Natl Acad Sci. 1931;17:656-660.

\bibitem{Palmer82} Palmer RG. Broken ergodicity. Adv Phys. 1982; 31:669-735.

\bibitem{Bantilan81} Bantilan FT Jr., Palmer RG. Magnetic-Properties of a Model Spin-Glass and the Failure of Linear Response Theory. J Phys F Met Phys. 1981; 11:261-266.

\bibitem{Reichl80} Reichl LE. A Modern Course in Statistical Physics. University of Texas Press, Austin (TX); 1980. 

\bibitem{Potts52} Potts RB. Some Generalized Order-Disorder Transformations. Proc Camb Phil Soc. 1952; 48:106-109.

\bibitem{Wu82} Wu FY. The Potts-Model. Rev Mod Phys. 1982; 54:235-268.

\bibitem{Gross84} Gross DJ, Mezard M. The Simplest Spin-Glass.  Nucl Phys B. 1984; 240:431-452.

\bibitem{Derrida80a} Derrida B. Random-Energy Model - Limit of a Family of Disordered Models. Phys Rev Lett. 1980; 45:79-82.

\bibitem{Derrida80b} Derrida B. The Random Energy-Model. Phys Rep. 1980;67:29-35.

\bibitem{Edwards75} Edwards SF, Anderson PW. Theory of Spin Glasses. J Phys F Met Phys. 1975; 5:965-974.

\bibitem{Janis05} Jani\v s V. Stability of solutions of the Sherrington-Kirkpatrick model with respect to replications of the phase space. Phys Rev B. 2005; 71:214403 (1-9).

\bibitem{Parisi80a} Parisi G. A Sequence of Approximated Solutions to the S-K Model for Spin-Glasses. J Phys A Math Gen. 1980; 13:L115-121. 

\bibitem{Parisi80b} Parisi G. Order Parameter for Spin-Glasses - Function on the Interval 0-1. J Phys A Math Gen. 1980; 13:1101-1112. 

\bibitem{Parisi80c} Parisi G. Magnetic-Properties of Spin-Glasses in a New Mean Field-Theory. J Phys A Math Gen. 1980; 13:1887-1895. 

\bibitem{Janis08} Jani\v s V. Free-energy functional for the Sherrington-Kirkpatrick model: The Parisi formula completed. Phys Rev B. 2008; 77:104417 (1-5).

\bibitem{Janis11b} Jani\v s V, Kl\'\i\v c A. Mean-field solution of the Potts glass near the transition temperature to the ordered phase. Phys Rev B. 2011; 84:064446 (1-10).

\bibitem{Janis13} Jani\v s V, Kauch A, Kl\'\i\v c A. Free energy of mean-field spin-glass models: Evolution operator and perturbation expansion. Phys Rev B. 2013;87:054201 (1-11).

\bibitem{Sherrington75} Sherrington D, Kirkpatrick S. Solvable Model of a Spin-Glass. Phys Rev Lett. 1975; 35:1792-1796.

\bibitem{Guerra03} Guerra F. Broken replica symmetry bounds in the mean field spin glass model. Commun Math Phys. 2003; 233:1-12.

\bibitem{Talagrand06} Talagrand M. The Parisi formula. Ann Math. 2006; 163:221-263.

\bibitem{Janis06} Jani\v s V, Kl\'\i\v c A. Hierarchical solutions of the Sherrington-Kirkpatrick model: Exact asymptotic behavior near the critical temperature. Phys Rev B. 2006; 74:054410 (1-9).                                                                                                                                                                      

\bibitem{Goldbart85} Goldbart P, Elderfield D. The Failure of the Parisi Scheme for Spin-Glass Models without Reflection Symmetry. J Phys C Solid State Phys. 1985; 18:L229-233.

\bibitem{Cwilich89} Cwilich G, Kirkpatrick TR. Mean-Field Theory and Fluctuations in Potts Spin-Glasses .1. J Phys A Math Gen. 1989; 22:4971-4987.

\bibitem{Gross85} Gross DJ, Kanter I, Sompolinsky H. Mean-Field Theory of the Potts Glass. Phys Rev Lett. 1985; 55:304-307.

\bibitem{Janis11a} Jani\v s V, Kl\'\i\v c A. Equilibrium state of the mean-field Potts glass. J Phys Condens Matter. 2011; 23:022204 (1-5).

\bibitem{Kirkpatrick87a} Kirkpatrick TR, Thirumalai D. Dynamics of the Structural Glass-Transition and the P-Spin-Interaction Spin-Glass Model. Phys RevLett. 1987; 58:2091-2094.

\bibitem{Kirkpatrick87b} Kirkpatrick TR, Thirumalai D. P-Spin-Interaction Spin-Glass Models - Connections with the Structural Glass Problem. Phys Rev B. 1987; 36:53-97. 
   
\end{thebibliography}
\end{document}